Biomarker Response to Galactic Cosmic Ray-Induced NO$_x$ and the Methane Greenhouse Effect in the Atmosphere of an Earthlike Planet Orbiting an M-Dwarf Star

John Lee Grenfell[1], Jean-Mathias Grießmeier[2*], Beate Patzer[3], Heike Rauer[1,3], Antigona Segura[4**], Anja Stadelmann[2], Barbara Stracke[1], Ruth Titz[1], and Philip von Paris[1]

(1) Institut für Planetenforschung
Extrasolare Planeten und Atmosphären
Deutsches Zentrum für Luft- und Raumfahrt (DLR)
Rutherford Str. 2
12489 Berlin
Germany

(2) Technische Universität Braunschweig
Mendelssohnstraße 3
38106 Braunschweig
Germany

(3) Zentrum für Astronomie & Astrophysik
Technische Universität Berlin (TUB)
Hardenbergstr. 36
10623 Berlin
Germany

(4) Department of Geosciences
443 Deike Building
Pennsylvania State University
University Park, PA 16802 USA

*Presently at: Laboratoire d'Etudes Spatiales et d'Instrumentation en Astrophysique (LESIA)
Centre National de la Recherche Scientifique (CNRS)-Observatoire de Paris
92195 Meudon
France

**Presently at: Infrared Processing and Analysis Center
California Institute of Technology
770 South Wilson Avenue
Pasadena, CA 91125, USA

*Abstract:* Planets orbiting in the habitable zone (HZ) of M-Dwarf stars are subject to high levels of galactic cosmic rays (GCRs) which produce nitrogen oxides in earthlike atmospheres. We investigate to what extent this $NO_x$ may modify biomarker compounds such as ozone ($O_3$) and nitrous oxide ($N_2O$), as well as related compounds such as water ($H_2O$) (essential for life) and methane ($CH_4$) (which has both abiotic and biotic sources) . Our model results suggest that such signals are robust, changing in the M-star world atmospheric column by up to 20% due to the GCR $NO_x$ effects compared to an M-star run without GCR effects and can therefore survive at least the effects of galactic cosmic rays. We have not however investigated stellar cosmic rays here. $CH_4$ levels are about 10 times higher than on the Earth related to a lowering in hydroxyl (OH) in response to changes in UV. The increase is less than reported in previous studies. This difference arose partly because we used different biogenic input. For example, we employed 23% lower $CH_4$ fluxes compared to those studies. Unlike on the Earth, relatively modest changes in these fluxes can lead to larger changes in the concentrations of biomarker and related species on the M-star world. We calculate a $CH_4$ greenhouse heating effect of up to 4K. $O_3$ photochemistry in terms of the smog mechanism and the catalytic loss cycles on the M-star world differs considerably compared with the Earth.

**1.1 Introduction**

We investigate the effect of Galactic Cosmic Rays (GCRs) upon atmospheric biomarkers ($O_3$ and $N_2O$) as well as $H_2O$ (essential to life) and $CH_4$ (which has biogenic and non-biogenic sources) (Des Marais et al., 2002) of an earthlike planet orbiting an active M-dwarf star in the Habitable Zone (HZ). We will henceforth refer to these compounds collectively as "biomarker and associated molecules". Such M-stars are important observational targets because (1) they are abundant in the solar neighbourhood (Tarter et al., 2006; Scalo et al., 2006, this issue) and (2) they have close-in HZs, favourable for transit observations of potentially habitable planets. The magnetosphere of a planet in the HZ of an M-star is probably much smaller than on Earth because: (a) the stellar wind flux is much higher because of the small orbital distance, and (b) the planetary magnetic dipole moment is reduced because the planetary rotation is limited by tidal locking (Grießmeier et al., 2005). Both these effects contribute to enhancing the flux of high energy cosmic ray particles into the planetary atmosphere. Scalo et al. (this issue, their section 4.2) provide an overview of the atmospheric chemistry of CRs, including the production of $NO_x$ and its influence upon biomarker and associated molecules.

In this contribution, we used GCR fluxes calculated from a magnetosphere model (Grießmeier et al., 2005) and we then adopted an air shower approach to calculate the corresponding atmospheric $NO_x$ source. Finally, we implemented this source into a photochemical column model to calculate the effect on biomarker and associated molecules. Section 2 describes the method/models and the runs made. Section 3 presents results, section 4 is the discussion and summary.

**1.2 Aim of this work**

Our main aim is to estimate the effect of high GCR fluxes upon biomarkers and associated molecules by implementing $NO_x$ sources from GCRs into our coupled atmospheric column model. Note that in this work we have investigated interplanetary GCR fluxes and interstellar GCR fluxes appropriate for our own solar system and just outside our own solar system, respectively. The work by Segura et al. (2005) was an important previous study, which used a different version of the model used in our work. The code differences between our work and Segura et al. (2005) are as follows:

(a) $CH_4$ and temperature coupling between the chemistry and climate routines has been improved. $N_2O$ coupling has been introduced.
(b) The weak lightning CO source in the chemistry routine has been removed.
(c) We employ the Jet Propulsion Laboratory (JPL) 2003 chemical kinetic data.
(e) After introducing effects (a) to (c), the surface biogenic fluxes were then retuned to reproduce Earth conditions in the Earth control. The new fluxes were applied to the M-star runs. As a result, we used methane surface fluxes of 735 teragramme (Tg) per year whereas the Segura study used 954 Tg per year. Observations for the Earth correspond to 600 Tg per year (Houghton et al. 1994). Differences arose due to missing processes (e.g. clouds) and/or missing chemistry (e.g. higher hydrocarbons which affects e.g. OH hence methane) in the model.

(f) The surface albedo was updated from 0.237 to 0.239 so that the mean surface temperature reproduced the Earth i.e. 288K.

Note that the Segura work considered both "active" and "quiet" M-stars whereas we consider only the former. The electromagnetic flux of "quiet" M-star drops sharply below 320nm compared with "active" M-stars. For M stars with no chromospheric activity, higher amounts of $N_2O$ and methane are expected due to the low UV emitted by these stars. In a different approach to our work, the Segura work chose to fix surface methane to an earthlike concentration in their radiative module for computational reasons. So, their chemistry module was allowed to calculate high methane values, but these values were not fed back into their radiation scheme.

## 2. Model Description and GCR parameterisation

**2.1 Cosmic Ray Proton Flux (CRPF) Model** - the flux of GCRs through the magnetospheres of different terrestrial exoplanets is calculated by Grießmeier et al. (2005, 2006). The number of protons of galactic origin reaching the top of a planetary atmosphere is calculated for the energy range 100 MeV< E<8 GeV. For each particle energy, 7 million particle trajectories are calculated. The planetary magnetosphere is assumed to be closed, and is modelled as a cylinder topped by a hemisphere of identical radius. This radius is determined by the pressure balance between the stellar wind ram pressure and the magnetic pressure of the planetary magnetic field. The CRPF calculates the particle fluxes (protons) at the top of the planetary atmosphere. Figure 1 shows some typical GCR flux spectra. The dashed line is for the top of the Earth's atmosphere output by the CRPF model. The thin, continuous line is for the top of our M-world atmosphere, again output by the CRPF model (Grießmeier et al. 2005, 2006). The medium thick line is for interplanetary space (Seo et al., 1994 ). This represents the upper limit for a planet without magnetospheric protection. The thickest line is for local interstellar space (Beer et al., 1991) i.e. this is the GCR flux without heliospheric/astrospheric protection by the stellar magnetic field.

**2.2 Cosmic Ray–$NO_x$ parameterisation** - GCR-induced $NO_x$ is formed via secondary electrons dissociating nitrogen molecules ($N_2$) followed by reaction of the resulting nitrogen atoms with atomic or molecular oxygen:

$$N_2 + e^- \rightarrow 2N + e^-$$
$$N + O_2 \rightarrow NO + O$$

We assumed the height-dependent $NO_x$ production rate to be:

$$P_{NO_x}(X) = k D_{N_2}(X) \qquad (1)$$

where $D_{N_2}$ is the destruction rate of molecular nitrogen and k the number of $NO_x$ produced per nitrogen molecule destroyed, which can be considered as a "quantum yield" and X is the overlying atmospheric mass. Since not only nitrogen atoms, but also ions are produced by the dissociation of the molecule, k must be less than 2. Various works have assumed different values for k (see, for example, Nicolet 1975: k = 0.96, Jackman et al. 1980: k = 1.25). Note there may be an altitude-dependence for k but it is not well-defined. Here we take k to be unity. $D_{N2}(X)$ is calculated as:

$$D_{N_2}(X) = \int_{E_1}^{E_2} n_{N_2}(X)\sigma_{total}(E_e)F_{el}(E_e)dE_e \qquad (2)$$

where $n_{N_2}$ is the number density of molecular nitrogen, $F_{el}$ the spectral electron flux, $\sigma_{total}$ the energy-dependent total destruction cross section and $E_1, E_2$ the limits of the energy range of electrons capable of dissociating molecular nitrogen. According to Nicolet 1975, the average cross section is $\sigma_{total} = 1.75 \ 10^{-16}$ cm² for electron energies between $E_1 = 30$eV and $E_2 = 300$eV.

To calculate the flux of secondary electrons, we adopted an air shower approach, which assumes that the incident protons produce electromagnetic cascades while travelling through the atmosphere. We assume a straight flight path in the atmosphere, thus neglecting any scattering. We can make the separation of the spectral electron flux $F_{el}$ into the total electron flux $R_{el}(X)$ and the spectral distribution of electrons, $S_{el}(E_e)$, so that $F_{el}(X, E_e) = R_{el}(X)S_{el}(E_e)$ with:

$$R_{el}(X) = \int_{\Omega} d\Omega \int_{Elow}^{E_{high}} N_{total}(X, E_p)dE_p \qquad (3)$$

where $\Omega$ is the solid angle, $E_{low}$ to $E_{high}$ represent the critical energy range for shower production. If the energy loss exceeds the proton energy from the CRPF model, no shower is generated. Above an energy of about 8 GeV the intensity of GCRs drops (Figure 1), so we set $E_{high} = 8.19$GeV. $N_{total}$ is the electron flux created by a proton of energy $E_p$ coming from a direction $(\theta, \varphi)$. Thus:

$$f = \int_{E_1}^{E_2} S_{el}(E_e) dE_e \quad (4)$$

where f is the fraction of electrons produced which can destroy $N_2$. We assume a third-power law for the normalized electron energy spectrum (Bichsel et al. 2005):

$$S_{el}(E_e) = aE_e^{-3} \quad (5)$$

where the constant, a is determined by the condition

$$\int_{E_{min}}^{E_{max}} S_{el}(E_e) dE_e = 1 \quad (6)$$

The respective boundaries are the minimal and the maximal energy of the electrons in the air showers. Assuming the air showers of different primary protons to be independent and identical to each other, and neglecting the dependence of the number of electrons produced, $N_{total}(X, E_p)$ upon proton energy, we can separate the electron flux into a directional number of electrons per shower, $N(X,\theta,\varphi)$, and the incoming angular proton flux at the top of the atmosphere $I_p$ as a function of proton energy: $N_{total}(X, E_p) = N(X,\theta,\varphi) I_p(E_p)$. This means that equation (3) results in:

$$R_{el}(X) = N_{el}(X) I_{int} \quad (7)$$

with

$$I_{int} = \int_{E_{low}}^{E_{high}} I_p(X=0, E_p) dE_p \quad (8)$$

and

$$N_{el}(X) = \int_\Omega N(X',\theta,\varphi) d\Omega \quad (9)$$

The proton energy spectrum, $I_p$ used is provided by the CRPF model (Grießmeier et al., 2005). $N(X',\Theta,\phi)$ describes the progress of the primary proton as it penetrates further into the atmosphere (Gaisser & Hillas 1977):

$$N(X') = N_{max} \left( \frac{X'-X_0}{X_{max}-X_0} \right)^{\frac{X_{max}-X_0}{\lambda}} e^{\frac{X_{max}-X'}{\lambda}} \quad (10)$$

where X' is the total atmospheric column mass density crossed by the primary particle, $N_{max}$ the number of particles at the shower maximum $X_{max}$, $X_0$ the height of the first interaction and $\lambda$ the attenuation length of the produced particles. X' is given by X' = $\frac{X}{\mu}$ with the atmospheric height measured by the mass column density X and $\mu = \cos\theta$ ($\theta$ incidence angle of the proton) as the usual definitions. Integrating over the solid angle in equation (9), assuming the atmosphere to be plane and the proton spectrum to be isotropic, gives:

$$N_{el}(X) = 2\pi \int_{\min(\frac{X}{X_0},1)}^{0} N_{max} \left(\frac{X'-X_0}{X_{max}-X_0}\right)^{\frac{X_{max}-X_0}{\lambda}} e^{\frac{X_{max}-X'}{\lambda}} d\mu \qquad (11)$$

This integral was evaluated numerically using a 5$^{th}$-order quadrature with 100 grid points. The calculated rate of $NO_x$ production from GCRs on Earth (see Figure 2, dashed line) reproduces the observed terrestrial values of Jackman et al. (1980) ( see their Figure 4) to within 10-20 %.

It is not the intention of this paper to calculate accurately the propagation of cosmic rays through the atmosphere, but to examine their eventual impact on atmospheric chemistry. Figure 2 shows our derived $NO_x$ production rates (molecules cm$^{-3}$ s$^{-1}$) calculated in our atmospheric column model for the same GCR fluxes as shown in Figure 1. The fluxes are set at the model's upper boundary (at around 70km). The calculation of the GCR-induced $NO_x$ rates were coupled with temperature and pressure profiles output by our atmospheric model. Figure 2 implies that the GCR-$NO_x$ production rates peak at 15-21km in the upper troposphere/lower stratosphere. The M-star case (thinnest, continuous line) leads to about three times more GCR-$NO_x$ production than for the Earth (dashed line). The thickest continuous line (GCR fluxes for local interstellar space) has $NO_x$ production about eight times higher than for the Earth. The resulting $NO_x$ production rates were implemented into the photochemical routine in the model.

**2.3 Column Model** - the original code has been described in detail by Kasting et al. (1984, 1985) and developed further by Segura et al. (2003). The climate module used is called the Rapid Radiation Transfer Module (RRTM), as employed by Segura et al. (2005). It uses the correlated k-distribution method of Mlawer et al. (1997) where 'k' is the absorption coefficient and employs 16 k-coefficients per wavelength interval with16 intervals between 3.1μm and $10^3$μm. This method groups together identical values of k in a spectral interval and calculates a mean absorption coefficient subsequently used in the transmittance calculation. Scattering is based on Toon et al. (1989). Note that although $CH_4$ features a pressure (Karkoschka, 1994) and temperature-dependence absorption in the near IR, this is currently not included in the model. Regarding the temperature-dependence, Sromovsky et al. (2006) suggested up to 40% increase in the near-IR absorption coefficient from 288K up to 330K (a typical surface temperature of the M-star runs). The incoming shortwave (SW) radiation routine in our model may require more k-values and an improved pressure parameterisation. These points will be explored in future work.

The chemistry includes 55 species for 220 reactions from the surface up to 64 km at 1 km intervals. The troposphere includes methane oxidation as well as wet and dry deposition. The chemistry features $HO_x$, $NO_x$, $O_x$ and $ClO_x$ families as well as their major reservoirs. Photolysis was diurnally-averaged for a cloud-free sky. For the runs described here we use the stellar spectrum of an M4.5V star, AD Leo, as described in Segura et al. (2005). Notice that AD Leo is a chromospherically active star, as a result it produces more UV radiation at wavelengths < 300 nm than the UV that will be emitted by an M star without an active chromosphere (Fig. 1B in Segura et al. 2005). For a discussion of how this may affect the abundance of biomarker and associated molecules in a planet orbiting such a star see the 'Discussion and Extensions' section in Segura et al. (2005).We chose the orbital distance of the planet (=0.16AU) so that the surface temperature, $T_s$, yields 288K. $CO_2$ was fixed to a modern-day Earth value of $3.55\times10^{-4}$ vmr. Note that our radiation scheme operates up to $CO_2$ levels of 3.5% volume mixing ratio. The chemistry levels extend up to 64km, the climate grid height is variable depending upon temperature and typically extends up to 70km for modern-day Earth conditions. Possible effects whereby GCRs deposit energy in the upper atmosphere are not included. We assumed terrestrial biota i.e. source emissions of $CH_4$, $N_2O$, CO and $CH_3Cl$ on the surface of the M-star planet were based on the Earth. Our work employed updated chemical kinetics based on the Jet Propulsion Laboratory (JPL) Report 2003. Appendix 1, Table A1, shows the main differences between the chemical kinetics used in this work compared with the Segura et al. (2003) work. The updated reactions differ by about 5-10% in their rates and the overall effect is not large. The photochemical model was integrated until the concentrations converged. Simulating a tidally-locked planet with a night and day face using a column model of the type employed here with averaged conditions is clearly a first approximation which depends on whether atmospheric density is sufficient to distribute quantities such as heat and momentum. We have performed some sensitivity studies for the Earth (not shown) which suggest that results are valid up to about 2 bar but at higher densities e.g. pressure-broadening effects and interpolation of the k-coefficients used in the climate code to derive the absorption coefficients make the results uncertain.

**2.4 About the Runs**

We performed in total five runs:

**Run (1) M-star run without methane-coupling** between chemistry and radiation. This run was performed to compare with Segura et al. (2005) who performed a similar calculation. Differences between the basic code used for this run in our work compared with that of the Segura work were discussed in section 1.2. Run 1 employed an M-dwarf (AD Leo) flux spectrum."Without methane coupling" means that $CH_4$ values used in the radiative transfer calculation were fixed at present day values corresponding to the Earth, whereas in the chemistry methane could build up to large values because OH concentrations were low, as already mentioned. High methane values in the chemistry were not passed to the climate subroutine.

**Run (2) M-star run with methane-coupling** – 'methane-coupling' means, changes in the $CH_4$ concentration calculated in the chemistry module are fed into the radiation module (in uncoupled mode, $CH_4$ in the radiation module was constant).

**Run (3) with top of atmosphere GCR induced NO$_x$ source** – as for run (2) but with GCR induced NO$_x$ sources (Figure 2, thin continuous line) introduced in the calculation of an earthlike planet orbiting an M-dwarf star in the HZ, as discussed above.

**Run (4) with interplanetary GCR induced NO$_x$ source** – as for run (2) but with interplanetary NO$_x$ sources (Figure 2, medium thick line). This represents a planet with a very weak magnetic moment, so that there is no magnetosphere shielding the planetary atmosphere against GCRs.

**Run (5) with interstellar GCR induced NO$_x$ source** – as for run (2) but with NO$_x$ sources (Figure 2, thickest line) derived from GCR fluxes for local interstellar space. This represents a star with a very weak magnetic field, so that there is no heliosphere/astrosphere shielding the planetary system against GCRs.

## 3. Results

### 3.1 Ozone comparison for this study with previous works

Table 2 shows a comparison of ozone column depths for various M-star exoplanet model scenarios:

| *Run* | *Total O₃ column (DU)* |
|---|---|
| *Segura et al. (2005)* <br> Earth control | 311 |
| *Our work* <br> Earth control | 294 |
| *Segura et al. (2005)* <br> AD Leo spectrum <br> No methane greenhouse | 164 |
| **Run 1** *(this work)* <br> AD Leo spectrum <br> No methane greenhouse | 681 |
| **Run 2** *(this work)* <br> As for run 1 but with methane greenhouse | 776 |
| **Run 3** *(this work)* <br> As for run 2 but with <br> M-world $NO_x$ from GCRs | 648 |
| **Run 4** *(this work)* <br> As for run 3 but with <br> interplanetary $NO_x$ from GCRs | 628 |
| **Run 5** *(this work)* <br> As for run 4 but with <br> interstellar $NO_x$ from GCRs | 548 |

Table 2: Ozone column depths shown in Dobson Units (DU) (1DU=2.7x10$^{16}$ molecules cm$^{-2}$) for various model scenarios.

In Table 2 for the Earth control we calculate a similar value (=294DU) as Segura et al. (2005) (=311DU) for the Earth control. For the AD Leo spectrum, however, we calculate widely different values. Segura et al. (2005) suggested a reduction by about 50% in column ozone (=164DU) for M-stars planets compared with solar planets (=311DU) which they associated with a weaker $O_3$ source due to slower $O_2$ photolysis for the M-star planet. Our study on the other hand, for similar input conditions, suggests that the ozone column more than doubles (e.g. 681DU, run 1) compared with the Earth. Ultimately, the discrepancy with the Segura work arose from relatively modest changes in the biogenic fluxes employed (see section 1.2 (e) above) . We could reproduce the Segura ozone value reasonably well when we adopted their biogenic fluxes, but changing these by quite modest values led to quite large changes in ozone.

Our high ozone values e.g. 681DU for Run 1, were mainly favoured by a slowing in the $HO_x$ and $NO_x$ cycles e.g. by factors of nine and eighteen compared with our Earth control run. Quantifying why the Segura work differed from this work would require a careful comparison of these catalytic cycles as part of a full source-sink analysis of ozone which is beyone the scope of this work. Our results however suggest, there may exist regimes in the M-world photochemistry, which depend sensitively on biospheric input and to which the ozone column may be particularly sensitive. The highly non-linear response between changing the biogenic flux by a modest amount and the ozone column response may indicate a strong positive feedback. The effect is not discerneable in the Earth control – it appears to be a facet of the M-star world photochemistry. As an example, we employ lower methane fluxes, compared with the Segura work, hence we calculate lower methane concentrations (discussed below) which influences OH hence has a wide-reaching influence on the chemistry.

Subsequent changes in the ozone column for Runs 2 to 5 in Table 2 are more modest. Evident is a gradual decrease with increasing $NO_x$ from GCRs which we interpret as an increasing role from ozone destruction via $NO_x$-based catalytic cycles. Table 2 nevertheless suggests that the ozone column can mostly survive the effects of GCRs. We now investigate the ozone photochemistry in more detail.

**Ozone response to GCRs** - On the Earth, 10% of the ozone column is produced by the so-called smog mechanism (Haagen-Smit et al. 1952) in the **troposphere**. This requires volatile organic compounds (e.g. methane), a $NO_x$ catalyst and sunlight. In the Earth's **stratosphere**, where 90% of the ozone column resides, oxygen photolysis produces ozone (Chapman chemistry) and catalytic cycles destroy ozone. So, GCRs can lead to ozone formation via the smog mechanism or to ozone removal via catalytic $NO_x$ cycles depending on which altitude the $NO_x$ is deposited. To estimate the relative importance of the two mechanisms (smog or Chapman), one can derive steady-state expressions which predict the concentration of ozone from smog alone or from Chapman alone.

For smog, the usual approach is to assume $NO_2$ is in steady-state i.e. loss $NO_2$ = production $NO_2$, assume the main loss is via photolysis and the main production is via reaction of NO with $O_3$:

$$\text{LOSS} = j_{NO2}[NO_2] = \text{PRODUCTION} = k_{NO+O3}[NO][O_3]$$

Re-arranging,   Smog $[O_3] = j_{NO2}[NO_2]/(k_{NO+O3}[NO])$ (1)

where $j_{NO2}$ is the photolysis coefficient for $NO_2$ ($s^{-1}$), $k_{NO+O3}$ is the rate coefficient for the reaction between NO and $O_3$, and square brackets represent concentrations in molecules $cm^{-3}$. For Chapman, the standard approach (e.g. Wayne, 1993) is to assume the Chapman reactions only, then to impose steady-state on $O_3+O^3P$, which leads to:

$$\text{Chapman } [O_3] = [O_2][(j_{O2}k_{(O+O3+M)}[M]/(k_{O3+O}j_{O3}))]^{0.5} \quad (2)$$

Where the symbols have similar meaning as in (1). Note that Chapman $[O_3]$ is actually an over-estimate because it predicts the ozone value from the Chapman cycle only and ignores ozone loss from the catalytic cycles. Comparing expressions (1) and (2) indicates which of the two mechanisms is dominating the ozone concentration in the model. Also, by comparing results from (1) and (2) with the actual ozone concentration calculated in the model stratosphere, we see how important the catalytic cycles are in lowering ozone compared with the Chapman value. We have calculated expressions (1) and (2) as diagnostics in our model. Table 3 shows their values (in $10^{-6}$ volume mixing ratio) for the various model runs at 30km, which corresponds to the heart of the ozone layer for the Earth's stratosphere:

| Run   | Model [$O_3$] | Smog [$O_3$] | Chapman [$O_3$] | %(Chapman-Model)/Chapman |
|-------|---------------|--------------|-----------------|--------------------------|
| Earth | 7.3           | 0.1          | 38.4            | 81                       |
| Run 1 | 15.8          | 9.6          | 45.2            | 65                       |
| Run 2 | 15.8          | 8.1          | 35.1            | 55                       |
| Run 3 | 14.3          | 7.7          | 41.6            | 66                       |
| Run 4 | 13.9          | 7.6          | 43.7            | 68                       |
| Run 5 | 12.2          | 7.1          | 53.3            | 77                       |

Table 3: 30km Ozone (in $10^{-6}$ vmr) calculated in the model ("Model $O_3$") compared with steady-state expressions indicating the $O_3$ concentration from the smog mechanism alone ("Smog $O_3$") and the Chapman mechanism alone ("Chapman $O_3$"). The final column is an indication of the %importance of the catalytic ozone cycles removing ozone.

Table 3 illustrates important differences between ozone photochemistry of the Earth compared with the M-star world. On the Earth run at 30km in the heart of the ozone layer, Chapman chemistry dominates the ozone production. For the M-star runs however, smog ozone takes on a more important role. For example, in run 1, the model calculates $15.8 \times 10^{-6}$ vmr ozone, and the smog expression alone suggests a value of $9.6 \times 10^{-6}$. The smog mechanism is important in the M-star runs because (1) methane levels are very high, as already discussed and (2) there is an extra $NO_x$ source from the GCRs. For runs 3 to 5, both (1) and (2) favour the smog production. Comparing the columns marked 'model $O_3$' and 'smog $O_3$' in Table 3, we see that the smog mechanism alone can account for about half of the ozone present in the model. The column marked 'Chapman $O_3$' in Table 3 features higher values for run 1 compared with the Earth, probably due to the differing UV environment of the M-star, as also discussed in Segura et al. (2005). In run 2, which undergoes some warming in the stratosphere due to the methane greenhouse effect, the Chapman chemistry slows – this is the expected negative-temperature dependence signal. For subsequent runs (runs three to five) the smog mechanism does not change greatly and Chapman increases. Despite this, model ozone decreases – the last column of Table 3 features increasing values, indicating an increasing role in the catalytic cycles. This is consistent with enhanced $NO_x$ from the GCRs with increasing run number. This implies that these cycles strengthen from runs three to five and contribute to the model ozone decrease.

Randall et al. (2005) noted that observed increases in CR $NO_x$ production of about a factor 4 led to about 60% ozone decrease in the Earth's atmosphere. However, our results differed from the Randall result, since our ozone was stable over quite a wide range of $NO_x$ production levels. A possible explanation for this discrepancy is, in the Earth's stratosphere, enhanced CR $NO_x$ leads mainly to catalytic loss (as observed by Randall's measurements) whereas on our M-star world, the smog mechanism (which is actually stimulated by enhanced $NO_x$) plays a more important role than on the Earth and ozone is not greatly depleted. Another reason for the difference between this work and Randall is, for the Earth CR $NO_x$ changes are large in the upper to middle atmosphere where ozone is abundant. In our results, most $NO_x$ is deposited lower down i.e. in the upper troposphere, due to the differing energy spectrum of the terrestrial compared with the M-star CRs. Note that stellar particle sources, not considered in this work, may also play a potentially significant role in affecting biomarkers and associated molecules. Solar Proton events, (SPEs) for example, could contribute a similar magnitude to stratospheric NO production as GCRs (Jackson et al., 1980, their Table 1) so our results should be viewed in this sense as a lower limit.

**3.2 Temperature Profiles**

Figure 3 shows temperature profiles for the runs 1 to 5. The methane greenhouse effect leads to a modest amount of extra heating through the column of about 2-4K. Including the GCRs (runs 3 to 5 in Figure 3) does not influence the temperature profile to a great extent. However, as we discussed, the methane level may be sensitive to small changes in the biogenic flux input. Values of up to $1 \times 10^{-3}$ vmr methane may be possible (Segura et al. (2005). Pavlov et al. (2000) suggested that hazes may form for $(CH_4/CO_2)>1$ which could lead to considerable cooling hence offset the methane greenhouse effect – such haze formation does not feature in our model.

**3.3 Column Amounts**

Figure 4 shows column values for the Earth (white) and for runs 1 to 5, shown in gradually darker shading. All column values increase for the M-star world, compared with the Earth. This was especially the case for methane and chloromethane, as already noted in previous works. The increase was mainly due to a sharp decline in OH, the main sink for these compounds, which decreased by around a factor of $10^5$ in the troposphere compared with the Earth control. Columns for a particular compound were all rather similar for runs one to five, suggesting that the signal can survive the effects of the GCRs, even for the extreme, interstellar scenario (run 5).

Although the biomarker and associated molecules survived in the runs presented here, note that $NO_x$ from the GCRs was mainly created around 20km. This is below the region where ozone peaks. Therefore it was difficult for the GCRs to impact ozone to a great extent. For e.g. more dense atmospheres, where the GCRs deposit their $NO_x$ at higher levels, if this region occurs close to the ozone maximum, then the GCRs will affect ozone to a larger extent. On Earth, we note that $NO_x$ cycles are important for regulating ozone in the lower to mid stratosphere. On our M-star world, however this may not be the case. For example, strong volcanic activity may increase the relative importance of $ClO_x$ cycles in the atmosphere. On the other hand, a somewhat warmer surface with a generous ocean coverage would likely stimulate atmospheric $HO_x$ cycle. Clearly, the $NO_x$ family must play an important role affecting ozone, if CRs are to play a major role.

**3.4 Profile Quantities**

Figure 5 shows altitude-concentration profiles for (a) ozone (b) methane (c) nitrous oxide and (d) water. The ozone peak has moved up to the upper stratosphere compared with around 30km for the Earth (not shown). Also, maximum values are higher by up to 50ppm. The higher values compared with the Earth were partly associated with a lowering in $HO_x$ and $NO_x$, as already discussed in section 3.1. Methane and nitrous oxide values change by only small amounts. The water profile increases up to about 40km above which it undergoes a sharp decrease. The onset of the decrease corresponds to the position of the cold trap (Figure 3) which has moved upwards compared with the Earth, where it occurs at 16-20km. The sharp decrease in water on the upper levels is consistent with the strong stellar flux in UV-B and EUV compared with the Earth.

Figure 6 shows the %changes in profiles due to GCRs for the same species as in Figure 5. For ozone (Figure 6a) we see a decrease of up to 25% from the surface up to 40km. It is initially puzzling that increasing $NO_x$ from GCRs (which, by itself should stimulate the smog mechanism) should lead to a lowering in ozone in a region where the smog mechanism operates. Further investigation implied that the reason was a drying effect (see Figure 6d) which led to a lowering in OH hence a slowing in the smog mechanism , since the first step of the mechanism involves reaction with OH. Methane and nitrous oxide changes were rather low in Figure 6. The water change (Figure 6d) is shown in absolute vmr units and not as a % due to very large changes in the % value in the upper layers, where absolute levels tended to zero. Evident in Figure 6d is the sensitivity of water to changes in the cold trap temperature at 40km, where a small warming (cooling) effects leads to modest increases (decreases) in water vapour. Note, for the upper levels in Figure 6d, zero values are not plotted.

## 4. Summary

- **Can biomarkers and associated molecules survive GCRs?**

Our results imply that GCR-induced $NO_x$ sources only modestly affects ozone and water concentrations and appear to be negligible for the other molecules considered. This suggests that biomarkers and associated compounds on these worlds are not destroyed by GCR-induced $NO_x$ chemistry, increasing the chances that they can be measured by forthcoming missions.

### Appendix 1

| Reaction | k (Segura et al. 2003) | k (this work) |
|---|---|---|
| $HO_2+O_3 \rightarrow OH+2O_2$ | $1.1 \times 10^{-14} exp(-500/T)$ | $1.0 \times 10^{-14} exp(-500/T)$ |
| $OH+O_3 \rightarrow HO_2+O_2$ | $1.6 \times 10^{-12} exp(-940/T)$ | $1.7 \times 10^{-12} exp(-940/T)$ |
| $HCO+O_2 \rightarrow HO_2+CO$ | $3.5 \times 10^{-12} exp(140/T)$ | $5.2 \times 10^{-12}$ |
| $H_2CO+OH \rightarrow H_2O+HCO$ | $1.0 \times 10^{-11}$ | $9 \times 10^{-12}$ |
| $H_2+O^1D \rightarrow OH+H$ | $1.0 \times 10^{-10}$ | $1.1 \times 10^{-10}$ |
| $CH_4+OH \rightarrow CH_3+H_2O$ | $2.9 \times 10^{-12} exp(-1820/T)$ | $2.45 \times 10^{-12} exp(-1775/T)$ |
| $CH_3O_2+HO_2 \rightarrow CH_3OOH+O_2$ | $3.8 \times 10^{-13} exp(800/T)$ | $4.1 \times 10^{-13} exp(750/T)$ |
| $CH_3O_2+NO \rightarrow H_3CO+NO_2$ | $4.2 \times 10^{-12} exp(180/T)$ | $2.8 \times 10^{-12} exp(300/T)$ |
| $N+NO \rightarrow N_2+O$ | $3.4 \times 10^{-11}$ | $2.1 \times 10^{-11} exp(100/T)$ |
| $NO+O_3 \rightarrow NO_2+O_2$ | $2.0 \times 10^{-12} exp(-1400/T)$ | $3.7 \times 10^{-12} exp(-1500/T)$ |
| $NO+HO_2 \rightarrow NO_2+OH$ | $3.7 \times 10^{-12} exp(250/T)$ | $3.5 \times 10^{-12} exp(250/T)$ |
| $NO_2+O \rightarrow NO+O_2$ | $6.5 \times 10^{-12} exp(120/T)$ | $5.6 \times 10^{-12} exp(180/T)$ |
| $NO_2+H \rightarrow NO+OH$ | $4.8 \times 10^{-10} exp(-340/T)$ | $4.0 \times 10^{-10} exp(-340/T)$ |
| $CH_3O_2+HO_2 \rightarrow CH_3OOH+O_2$ | $3.8 \times 10^{-13} exp(800/T)$ | $4.1 \times 10^{-13} exp(750/T)$ |
| $CH_3O_2+HO_2 \rightarrow CH_3OOH+O_2$ | $3.8 \times 10^{-13} exp(800/T)$ | $4.1 \times 10^{-13} exp(750/T)$ |

**Table A1:** Differences in the chemical kinetic data used in the Segura et al. (2003) work compared with the present work, which adopted data from DeMore et al, (2003). Units are k (molecules$^{-1}$ cm$^3$ s) and T (Kelvin).

**Figure Captions**

Figure 1: Galactic Cosmic Ray (GCR) Fluxes for (a) the top of atmosphere for the Earth (dashed line) (b) the top of atmosphere for an earthlike planet orbiting in the HZ of an active M-dwarf at 0.16AU (thin continuous line) (c) interplanetary solar-system space (medium continuous line) and (d) local interstellar space (thick continuous line).

Figure 2: Rate of NO production (molecules cm$^{-3}$ s$^{-1}$) calculated in an atmospheric column model whose upper boundary was subjected to the four GCR flux scenarios shown in Figure 1.

Figure 3: Modelled temperature (K) for (a) run 1 (plain line); run 2 (long dashed line); run 3 (short dashed line); run 4 (long-short dashed line) and run 5 (dotted line).

Figure 4: Modelled column values shown in Dobson Units (DU). Earth values are shown in white. Runs 1 to 5 are depicted in gradually darker shading where run 1=lightest grey and run 5=black.

Figure 5: Concentration profiles of (a) ozone (b) methane (c) nitrous oxide and (d) water for run 1 (plain line); run 2 (long dashed line); run 3 (short dashed line); run 4 (long-short dashed line) and run 5 (dotted line).

Figure 6: As for Figure 5 but showing changes due to changing GCR-induced NO$_x$ sources. Results are shown relative to run 2 which is without GCR sources. Plotted is the value ((run x- run 2) / run 2) *100% where x=run 3 (long dashed line), run 4 (short dashed line) and run 5 (dotted line), except for water, which shows the difference (run with GCRs – run 2 without GCRs) in 10$^{-6}$ vmr. The water results (Figure 6d) are not plotted above 50km to avoid very large values which arose because the denominator value approached zero.

**Acknowledgements**

We are grateful to two anonymous referees for helpful comments. We are also grateful to Dr. James Kasting for providing original code and for useful discussion. Dr Eli Mlawer provided useful advice and help regarding the radiative routine employed. This study was supported by the International Space Science Institute (ISSI) and benefited from the ISSI Team ``Evolution of Habitable Planets''. A. S. thanks the NASA Astrobiology Institutes Virtual Planetary Laboratory Lead Team, supported by the National Aeronautics and Space Administration through the NASA Astrobiology Institute under Cooperative Agreement Number CAN-00- OSS-01.